\documentclass[preprint, 12pt, authoryear, aas_macros]{elsarticle}

\usepackage{amssymb}

\journal{Icarus}

\begin{document}

\begin{frontmatter}



\title{Zonal Flow Magnetic Field Interaction in the Semi-Conducting Region of Giant Planets}


\author{Hao Cao and David J. Stevenson}

\address{Division of Geological and Planetary Sciences, California Institute of Technology, Pasadena, CA 91125}

\begin{abstract}

All four giant planets in the Solar System feature zonal flows on the order of 100 $m/s$ in the cloud deck, and large-scale intrinsic magnetic fields on the order of 1 $Gauss$ near the surface. The vertical structure of the zonal flows remains obscure. The end-member scenarios are shallow flows confined in the radiative atmosphere and deep flows throughout the entire planet. The electrical conductivity increases rapidly yet smoothly as a function of depth inside Jupiter and Saturn. Deep zonal flows will inevitably interact with the magnetic field, at depth with even modest electrical conductivity. Here we investigate the interaction between zonal flows and magnetic fields in the semi-conducting region of giant planets. Employing mean-field electrodynamics, we show that the interaction will generate detectable poloidal magnetic field perturbations spatially correlated with the deep zonal flows. Assuming the peak amplitude of the dynamo $\alpha$-effect to be 0.1 $mm/s$, deep zonal flows on the order of 0.1 -- 1 $m/s$ in the semi-conducting region of Jupiter and Saturn would generate poloidal magnetic perturbations on the order of 0.01 \% -- 1 \% of the background dipole field. These poloidal perturbations should be detectable with the in-situ magnetic field measurements from the Juno mission and the Cassini Grand Finale. This implies that magnetic field measurements can be employed to constrain the properties of deep zonal flows in the semi-conducting region of giant planets.

\end{abstract}

\begin{keyword}

 Jupiter \sep Saturn \sep Interiors \sep Magnetic Fields



\end{keyword}

\end{frontmatter}


\section{Introduction}
\label{}

The giant planets in the solar system are natural laboratories for rotating convection and magnetohydrodynamics. The existence of deep convection inside all four giant planets are guaranteed by the measured large intrinsic heat flux (``large'' here means larger than the heat flux that can be conducted along the adiabats) and the large-scale intrinsic magnetic fields. The dynamical details of the deep convection (e.g. amplitude, structure, and energy partitioning) and the coupling to the shallow atmospheric dynamics, however, remain largely unknown. In terms of observations, the gravity and magnetic field measurements from the Juno mission \citep[]{Bolton2010} and the Cassini Grand Finale \citep[]{Spilker2014} will provide an unprecedented opportunity to constrain the interior dynamics of Jupiter and Saturn.  



\begin{figure}[h!]
 \centering
     \includegraphics[width=0.95\textwidth]{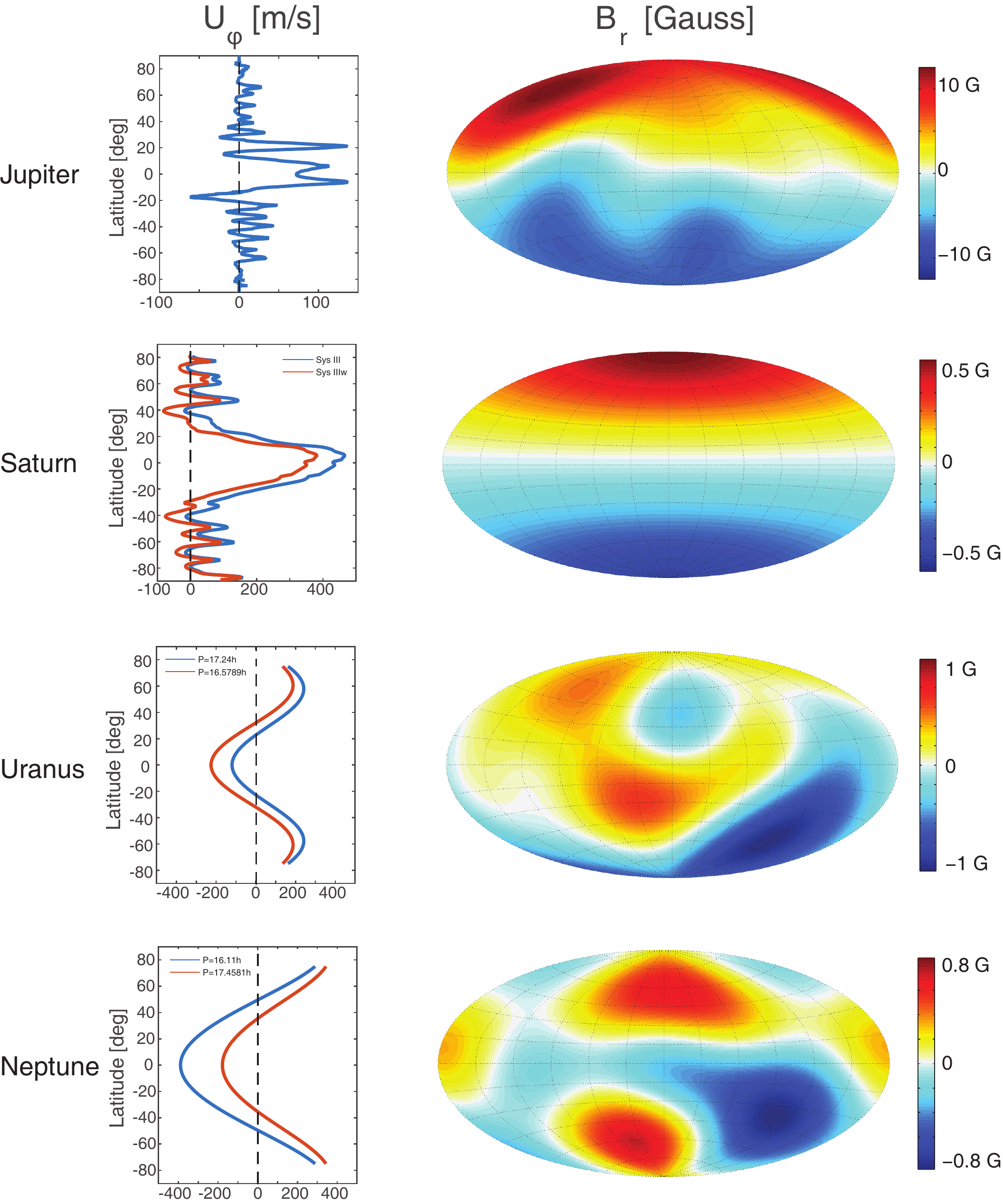}
 \caption{Observed surface zonal wind and magnetic field profile for solar system giant planets. The zonal wind profile for Jupiter and Saturn shown here are from Cassini and Voyager observaions \citep[]{Porco2003, SL2000, Vasavada2006, Baines2009}, while the zonal wind profile for Uranus and Neptune shown here are the empirical fits to Hubble Space Telescope (HST) and Voyager 2 observations \citep[]{Sromovsky1993, Sromovsky2001, Sromovsky2005, Hammel2005}. Not many observational constraints exist for zonal winds on Uranus and Neptune at latitude beyond 75 degrees north and south, but the winds likely decrease to zero smoothly towards the poles. The uncertainties in our understanding of the deep interior rotation rates of Saturn, Uranus and Neptune affect the details of the surface zonal wind profiles. The surface magnetic field profiles are based on the Galileo 13 model for Jupiter \citep[]{Yu2010}, Cassini 5 model for Saturn \citep[]{Cao2012}, Umoh model and Nmoh model truncated at degree and order 3 for Uranus and Neptune \citep[]{HB1996}.}
 \label{fig:fig1}
\end{figure}

Dynamics in the atmospheres of the solar system giant planets have been inferred from cloud tracking \citep[]{Porco2003, SL2000, Vasavada2006, Baines2009, Sromovsky1993, Sromovsky2001, Sromovsky2005, Hammel2005}. The dominant features of the atmospheric dynamics of all four giant planets are the east-west zonal winds on the order of 100 $m/s$ (Fig. \ref{fig:fig1}). In the equatorial region, Jupiter and Saturn feature super-rotation (eastward wind in the corotation frame), while Uranus and Neptune feature sub-rotation (westward wind in the corotation frame). Jupiter's off-equatorial region features zonal winds with alternating directions, with the eastward flows being stronger than the westward flows when viewed in the System III corotation frame. Saturn's off-equatorial region features zonal flows with varying speeds. The few minutes uncertainties in our understanding of Saturn's deep interior rotation rate translate into uncertainties in the direction of the off-equatorial winds as well as the width of the equatorial super-rotation (Fig. \ref{fig:fig1}). The off-equatorial region of Uranus and Neptune feature one broad sub-rotation in each hemisphere. The latitude of transition from sub-rotation to super-rotation on the surfaces of Uranus and Neptune are affected by the uncertainties in our understanding of the deep interior rotation rates (Fig. \ref{fig:fig1}). 

Intrinsic magnetic fields have been detected for all four giant planets \citep[]{Connerney1993, Connerney2007, Cao2011, Cao2012}. In terms of amplitude, the surface magnetic fields of Saturn, Uranus and Neptune are about 0.3 $Gauss$ (30,000 $nT$), while the surface magnetic field of Jupiter is about 6 $Gauss$ (600,000 $nT$). In terms of morphology, the magnetic fields of Jupiter and Saturn are axial dipole dominant, while the magnetic fields of Uranus and Neptune are non-axial and multipolar. Jupiter's magnetic dipole axis is tilted about 10 degrees from the spin axis, while Saturn's magnetic dipole axis is aligned with the spin-axis to within 0.06 degrees according to the latest Cassini measurements \citep[]{Cao2011}. 


Electrical conductivity inside Jupiter and Saturn increases rapidly yet smoothly from $< 10^{-7}$ $S/m$ near the 1 bar level to $10^5 - 10^6$ $S/m$ near the $1 - 3$ Mbar level \citep[]{Weir1996, Nellis1996, Liu2008, French2012}. The high electrical conductivity in the deep interior is likely due to pressure ionization of hydrogen. The alkali metals with solar composition would be the main contributor to the electrical conductivity in the low pressure region. Electrical conductivity inside Uranus and Neptune is uncertain due to two unknowns: the abundance of ice (water, methane, ammonia) in the hydrogen-helium envelope and the abundance of hydrogen in the ice layer. The electrical conductivity inside ice giant would only reach $10^3$ $S/m$ in the ice layer without significant mixing of hydrogen \citep[]{Nellis1997}, and would remain below $1$ $S/m$ in the hydrogen-helium envelope if the ice mixing ratio in the envelope is below 10\% \citep[]{Liu2006}.  


The highly conducting region of giant planets with electrical conductivity greater than 1000 $S/m$ likely feature zonal flows on the order of 1 $cm/s$ or less, based on Jovian magnetic secular variation measurements \citep[]{Yu2010, RH2016}. The magnetic secular variation measurements are not straightforward to interpret for giant planets for two reasons. First, the rotation rate of the deep interior of giant planets is defined by the observed rotation rate of non-axisymmetric magnetic fields. Thus only the spatial variation of the drifts in the non-axisymmetric magnetic fields (e.g., caused by spatial variation of steady deep zonal flow) and the time variations of the drifts (e.g., caused by the time variations of deep zonal flow) can be straightfowardly detected. Second, the forward problem of observable magnetic secular variations for a given deep flow structure has not been solved for a planet with radially varying electrical conductivity. We will address the second problem in detail in a separate paper. For now, we interpret the inferred Jovian magnetic secular variation loosely as representing flows in regions with magnetic Reynolds number $(Rm)$ greater than 10. We choose $Rm=10$ as the threshold to ensure the frozen-in condition is satisfied. This interpretation indicates that deep zonal flows inside Jupiter with magnetic Reynolds number greater than 10 should be on the order of 1 $cm/s$ or less. The case for Saturn is less clear. Neither any departure from axisymmetric nor any secular variation in the axisymmetric magnetic field has been detected \citep[]{Cao2011}, which indicates the possibility of stronger zonal flows in the highly conducting region of Saturn. In fact, a relatively strong zonal-dominant flow in the outer part of the highly conducting region of Saturn could be a natural explanation for the extreme axisymmetry of the observed intrinsic magnetic field \citep[]{Stevenson1980, Stevenson1982}. A stable compositional gradient (e.g. set up by helium rain out from hydrogen in the Mbar region) could help ensure the dominance of zonal flows in the outer part of the highly conducting region of Saturn \citep[]{Stevenson1980, Morales2009, Lorenzen2009, Wilson2010}. 

The observed intrinsic heat flow of giant planets set an upper bound on the internal Ohmic dissipation, when viewed on long time scales. The calculation of Ohmic dissipation is robust and straightforward for regions with low magnetic Reynolds number (e.g. $Rm < 1$), since the magnetic field there can still be robustly estimated from potential field continuation. The Ohmic dissipation constraint excludes 100 $m/s$ zonal winds in regions with electrical conductivity higher than 0.01 $S/m$ for Jupiter and Saturn \citep[]{Liu2008}. However, zonal winds on the order of $1$ $m/s$ could well reside in the semiconducting region of giant planets. Detection of zonal winds on the order of 1 $m/s$ in the semiconducting region of giant planets would indicate a relatively smooth transition of zonal flows from surface to the deep interior. Strictly speaking, the internal Ohmic dissipation can exceed the observed surface luminosity by a factor of $(T_{Interior}/T_{Surf}-1)$ \citep[]{Backus1975, Hewitt1975, Jones2008}. Here $T_{Interior}$ and $T_{Surf}$ are the temperatures in the deep interior and temperature near the surface of the convective envelope respectively. This ratio is around 40 for Jupiter when considering dissipation outside 0.90 $R_J$. However, given the super exponential increase of the electrical conductivity as a function of depth inside Jupiter and Saturn, a factor of 40 makes little difference in terms of constraining the depth of deep zonal flows from Ohmic dissipation considerations \citep[]{Liu2008}.

 Three-dimensional magnetohydrodynamic (MHD) simulations with strong radial variation of electrical conductivity and density have been carried out \citep[]{Jones2014, Gastine2014, Duarte2013}. Limited by the currently available computational power, relatively high viscosity needs to be adopted in these simulations. A common finding from these simulations is that dipole-dominant magnetic field and strong deep zonal flows are incompatible. A single band of equatorial super-rotation confined in the low electrical conductivity region are the most common features in the solutions with a dipole-dominant magnetic field. Given the difficulty of obtaining off-equatorial zonal jet in the currently accessible parameter space of deep shell numerical MHD simulations, the interaction between off-equatorial jet and deep dynamo generated magnetic field likely needs to be addressed in reduced models.

In this paper, we investigate the inevitable interaction between zonal flows and magnetic fields in the semi-conducting region of giant planets. In particular, we will show that such interactions, in the background of rotating turbulent convection, will generate detectable poloidal magnetic field perturbations. In section 2, we present a qualitative description of the physics of zonal flow magnetic field interactions in the semi-conducting region of giant planets. The equation of mean-field electrodynamics and its simplification under the small poloidal perturbation limit are presented in section 3. Section 4 and 5 present our analyses and calculations for Jupiter and Saturn, followed by discussion and summary in section 6 and 7 respectively.

\section{The Physics of Zonal Flow Magnetic Field Interaction in the Semi-Conducting Region of Giant Planets}
\label{}

\begin{figure}[h!]
 \centering
     \includegraphics[width=0.95\textwidth]{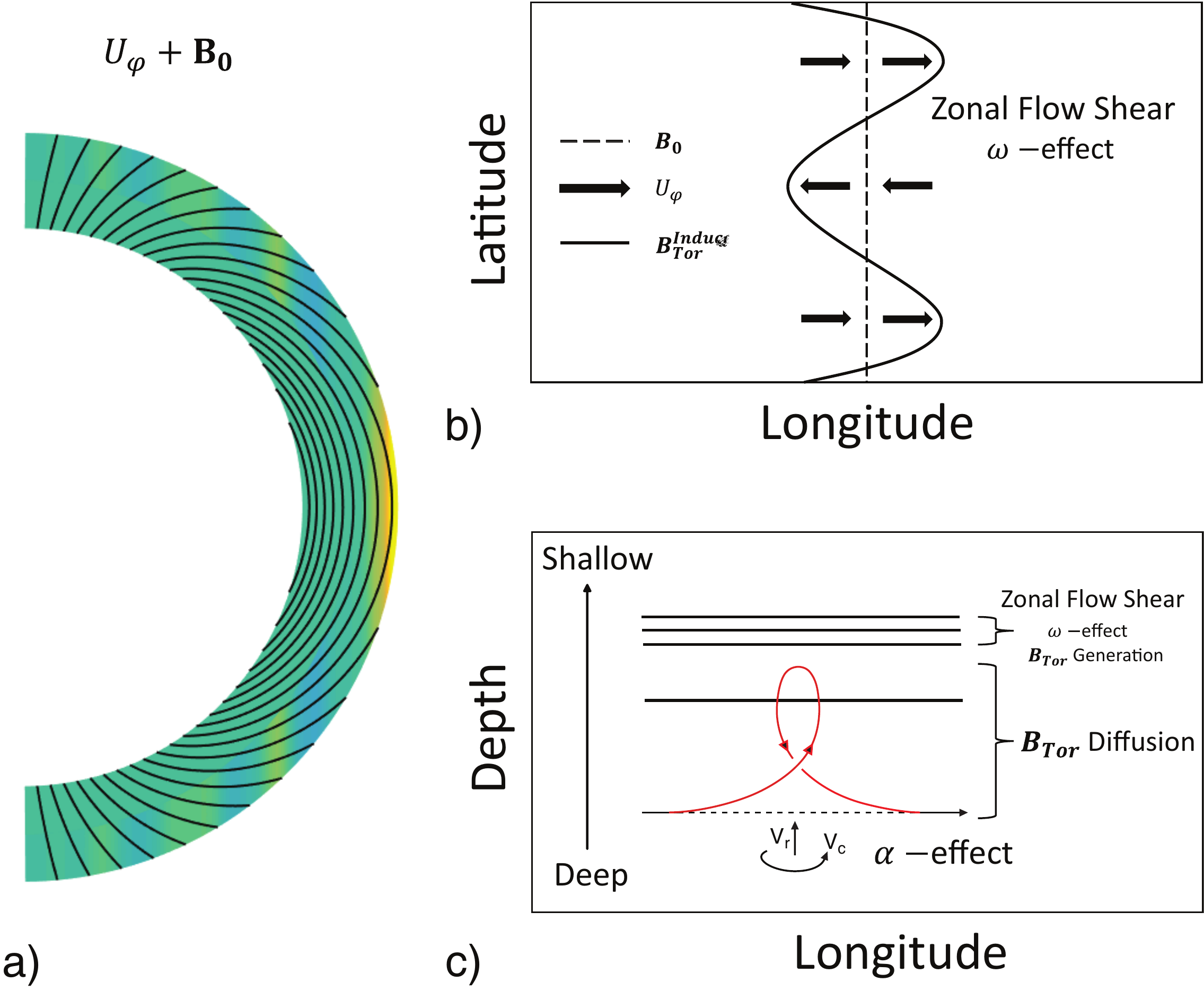}
 \caption{Physics of zonal flow magnetic field interaction in the semi-conducting region of giant planets. The interaction consists of two ingredients: the generation of toroidal magnetic field by the zonal flows acting on the background poloidal magnetic field (the dynamo $\omega$-effect), and the generation of poloidal magnetic perturbations by the small-scale turbulent convection acting on the toroidal magnetic field (the dynamo $\alpha$-effect). Both the toroidal and the poloidal magnetic field resulted from this interaction would be spatially correlated with the deep zonal flows. Due to the relatively low electrical conductivity in the semi-conducting region, spatially separated $\omega-$effect and $\alpha-$effect can communicate effectively through magnetic diffusion. Toroidal magnetic field generated by zonal flows at relatively shallow depth can diffuse downward to deeper regions where the $\alpha-$effect is more efficient (panel c).}
 \label{fig:fig2}
\end{figure}

Two physical features uniquely define the physics of zonal flow magnetic field interaction in the semi-conducting region of giant planets. The first is the rapidly yet smoothly radial-varying electrical conductivity, and the second is the presence of a deep dynamo (Figs. \ref{fig:fig2} \& \ref{fig:fig3}). Smoothly radial-varying electrical conductivity indicates a smooth transition from hydrodynamics to magnetohydrodynamics (MHD) inside Jupiter and Saturn. The electric current and Lorentz force would increase smoothly as a function of depth. The presence of a deep dynamo imposes a large-scale magnetic field in the semi-conducting region. This background large-scale magnetic field is anchored to the $\sim 1$ $cm/s$ flow in the deep dynamo region. Zonal flow, meridional circulation, and turbulent convection in the semi-conducting region will modify the background magnetic field via shear, advection, stretch and twist. The Lorentz force associated with these actions will back-act on the flows. The back-action of Lorentz force likely enters the dynamical balance of flows inside giant planets. However, the role of Lorentz force on planetary interior flows is still poorly understood. 

At present, we restrict our attention to the kinematic problem: the modification of the dynamo generated magnetic field by zonal flows in the semi-conducting region. The full dynamical problem is yet to be solved. Here we qualitatively describe the physics of the kinematic interaction. The mathematical details will be given in the next section. The interaction consists of two ingredients (Fig. 2): the generation of toroidal magnetic field by the zonal flows acting on the background poloidal magnetic field (the dynamo $\omega$-effect), and the generation of poloidal magnetic perturbations by the small-scale turbulent convection acting on the toroidal magnetic field (the dynamo $\alpha$-effect). Both the toroidal and the poloidal magnetic fields resulted from this interaction would be spatially correlated with the deep zonal flows. 

The interaction between flow and magnetic field in the semi-conducting region is distinct from that in the deep dynamo region in several ways. First of all, a crucial distinction between the two concerns the validity of our current analytical tools to deal with the problem. The first order smoothing approximation (FOSA) in the derivation of the dynamo $\alpha-$effect is strictly valid in the semi-conducting region where the magnetic Reynolds number associated with small-scale flow is small. The validity of the dynamo $\alpha-$effect in describing the magnetic field generation process in the deep dynamo region, where Rm associated with small-scale flow is large, has been seriously challenged \citep[]{Boldyrev2005, Cattaneo2006, Cattaneo2009, HughesProctor2009, Cattaneo2014}. Second, due to the relatively low electrical conductivity in the semi-conducting region, magnetic diffusion is more pronounced. As a result, spatially separated $\omega-$effect and $\alpha-$effect can communicate effectively through magnetic diffusion (Fig. 2c). The electrical conductivity gradient dictates an asymmetric magnetic diffusion: downward diffusion will be more pronounced than upward diffusion due to higher electrical conductivity deep down. Toroidal magnetic field generated by zonal flows at relatively shallow depth can diffuse downward to deeper regions where the $\alpha-$effect is more efficient. Thus the poloidal field generated by the interaction will likely exceed the estimations from local approximations. Third, the interaction is only expected to produce a small modification to the pre-existing magnetic field originated from the deep dynamo region. Energetic consideration strongly favors the wind induced poloidal magnetic field in the semi-conducting region being small compared to the deep dynamo generated magnetic field. Compare to the semi-conducting region, it is several orders of magnitude more efficient to drive the same amount of electric current in the deep dynamo region due to the several orders of magnitude higher electrical conductivity. Moreover, if the poloidal field generated by the interaction in the semi-conducting region are comparable to the deep dynamo generated field, the magnetic field is expected to become oscillatory as shown in the classical work of $\alpha-\omega$ dynamo by \citet[]{Parker1955} and fully dynamical MHD simulations \citep[e.g.][]{Brown2011, Kapyla2012, Gastine2012, Dietrich2013, Augustson2015}. Given the extremely tight upper bound on the time variation of the dipole moment of Jupiter and Saturn \citep[]{Yu2010, RH2016, Cao2011}, an oscillatory magnetic field with a period on the order of $\sim$ 100 $years$ or shorter can be safely ruled out. These considerations indicate that the poloidal magnetic field generated by the flows in the semi-conducting region likely is a small perturbation to the deep dynamo generated poloidal magnetic field. This realization would enable us to simplify the dynamo equations in the semi-conducting region considerably. 

\section{The Mean-Field Electrodynamics Equation under the Small \\Poloidal Perturbation Limit}
\label{}

To quantitatively describe zonal flow magnetic field interaction in the semi-conducting region of giant planet, we turn to mean-field electrodynamics \citep[]{Moffatt1978, KR1980}. The essential step in the development of mean-field electrodynamics is the closure of the mean-field equation: the time evolution of the mean magnetic field depends only explicitly on the mean flow and the mean magnetic field, while the interaction between the small-scale flow and the small-scale magnetic field can be effectively described by the dynamo $\alpha-$effect. This closure is not guaranteed in general. Under the condition where the magnetic Reynolds number associated with the small-scale flow ($Rm'$) is smaller than unity, this closure can be achieved. Under this condition, the small-scale magnetic fields owe their existence entirely to the small-scale flow acting on the mean magnetic field. As one moves from the deep interior towards the outer part of the giant planets with relatively low electrical conductivity, there will be a cut-off radius beyond which $Rm'$ is smaller than unity. This cut-off radius sets the lower boundary for the semi-conducting region of giant planets, which ensures the validity of mean-field electrodynamics in this part of the giant planets. Any attempt to apply mean-field electrodynamics to the deep dynamo region of giant planets should be cautioned. After a brief introduction of the mean-field electrodynamics equation, we derive the simplification of it under the small poloidal perturbation limit.  

\subsection{The Mean-Field Electrodynamics Equation}

In mean-field electrodynamics (Moffatt, 1978; Krause \& Radler, 1980), the governing equation for the mean magnetic field $\overline{\mathbf{B}}$ reads
\begin{equation} 
   \frac{\partial \overline{\mathbf{B}}} {\partial t} = \nabla \times (\overline{\mathbf{U}} \times \overline{\mathbf{{B}}}  +  \alpha  \overline{\mathbf{{B}}} - \eta_E \nabla \times  \overline{\mathbf{{B}}}), \label{eq:MFED}
\end{equation}
where $\eta_E$ is the effective magnetic diffusivity
\begin{equation}
\eta_E=\eta+\beta.
\end{equation}
Here $\eta$ is the magnetic diffusivity ($\eta=1/\mu_0\sigma$, where $\mu_0$ is the magnetic permeability of free-space and $\sigma$ is the electrical conductivity), $\alpha$ is the dynamo $\alpha-$effect, and $\beta$ is the turbulent diffusivity, and they can be estimated as
\begin{equation}
\alpha \sim f l'u'^2/\eta,
\end{equation}
\begin{equation}
\beta \sim u'l',
\end{equation}
under FOSA where $l'$ is the typical length scale of the turbulent convective cells, $f$ is a coefficient measuring the relative kinetic helicity of the flow 
\begin{equation}
f=-\frac{\overline{\mathbf{u'}\cdot (\nabla \times \mathbf{u'})}}{\overline{|u'||\nabla \times \mathbf{u'}|}}.
\end{equation}
It should be noted that $\alpha$ takes the unit of velocity and $\beta$ takes the unit of diffusivity. From here on, we work with the mean-fields only and drop the over-line in the symbols for simplicity.


\subsection{The Small Poloidal Perturbation Limit}

As discussed in section 2, the following conditions are likely true for zonal flow magnetic field interaction in the semi-conducting region of giant planets. 1) The dominant axial dipole components observed at Jupiter and Saturn originate from the deep dynamo region. 2) The poloidal magnetic field resulted from the interaction among zonal flow, turbulent convective flow, and the background dipolar magnetic field in the semi-conducting region is small compared to the background magnetic field. 3) The toroidal magnetic field resulted from the interactions, on the other hand, needs not to be small compared to the background magnetic field.

Under these conditions, the mean-field equation under axisymmetric decomposition $\mathbf{B}(r,\theta)=\nabla \times (A\hat{\mathbf{e_\phi}}) + B \hat{\mathbf{e_\phi}}$ gets simplified to
\begin{equation}
\frac{\partial A}{\partial t} + \frac{1}{s} \mathbf{U_P}\cdot \nabla (s A)= \alpha B + \eta_E (\nabla^2 - \frac{1}{s^2})A,
\end{equation}
\begin{equation}
\frac{\partial B}{\partial t} + s \nabla \cdot (\frac{B}{s} \mathbf{U_P})= s\mathbf{B_0}\cdot \nabla \omega + \eta_E (\nabla^2 - \frac{1}{s^2})B + \frac{1}{r}\frac{d\eta_E}{dr}\frac{\partial (rB)}{\partial r}, 
\end{equation}
where $\mathbf{B_0}$ is the background planetary dipolar magnetic field, $s=r \sin \theta$, $\mathbf{U_P}$ is the meridional circulation, and $\omega$ is the angular velocity $\omega={U_\varphi}/{s}$. The above two equations are partially decoupled: the poloidal potential $A$ depends on the toroidal magnetic field $B$, however, $B$ is independent of $A$.

It can be shown that the magnetic Reynolds number of the meridional circulation is much smaller than unity in the semi-conducting region (see Appendix D). Thus, the meridional circulation can be ignored as a first step, the mean-field equations get further simplified to
\begin{equation}
\frac{\partial A}{\partial t} = \alpha B + \eta_E (\nabla^2 - \frac{1}{s^2})A,
\label{eqn:A_Final}
\end{equation}
\begin{equation}
\frac{\partial B}{\partial t} = s\mathbf{B_0}\cdot \nabla \omega + \eta_E (\nabla^2 - \frac{1}{s^2})B + \frac{1}{r}\frac{d\eta_E}{dr}\frac{\partial (rB)}{\partial r}.
\label{eqn:B_Final}
\end{equation}
The steady-state solution to the above two equations are then sought through spectral decompositions. The spectral representation of the above equations and some details of the numerics are given in Appendix A.  

The steady-state solution to partially decoupled mean-field equations (\ref{eqn:A_Final} - \ref{eqn:B_Final}) are further compared with time-stepping of the fully coupled mean-field equations (\ref{eqn:A_F2} - \ref{eqn:B_F2} in Appendix B). The steady-state solution and the time-stepping solution agree to within 5\% when the wind-induced poloidal perturbations are smaller than 20\% of the background field (see Appendix B). The simplicity of the steady-state solution, and the simple physical picture underlying it, makes it the preferred way to obtain solution for this problem.

\section{Order of Magnitude Analysis for Jupiter and Saturn}

\subsection{The Definition of the Semi-Conducting Region of Jupiter and Saturn}

\begin{figure}[h!]
 \centering
     \includegraphics[width=0.95\textwidth]{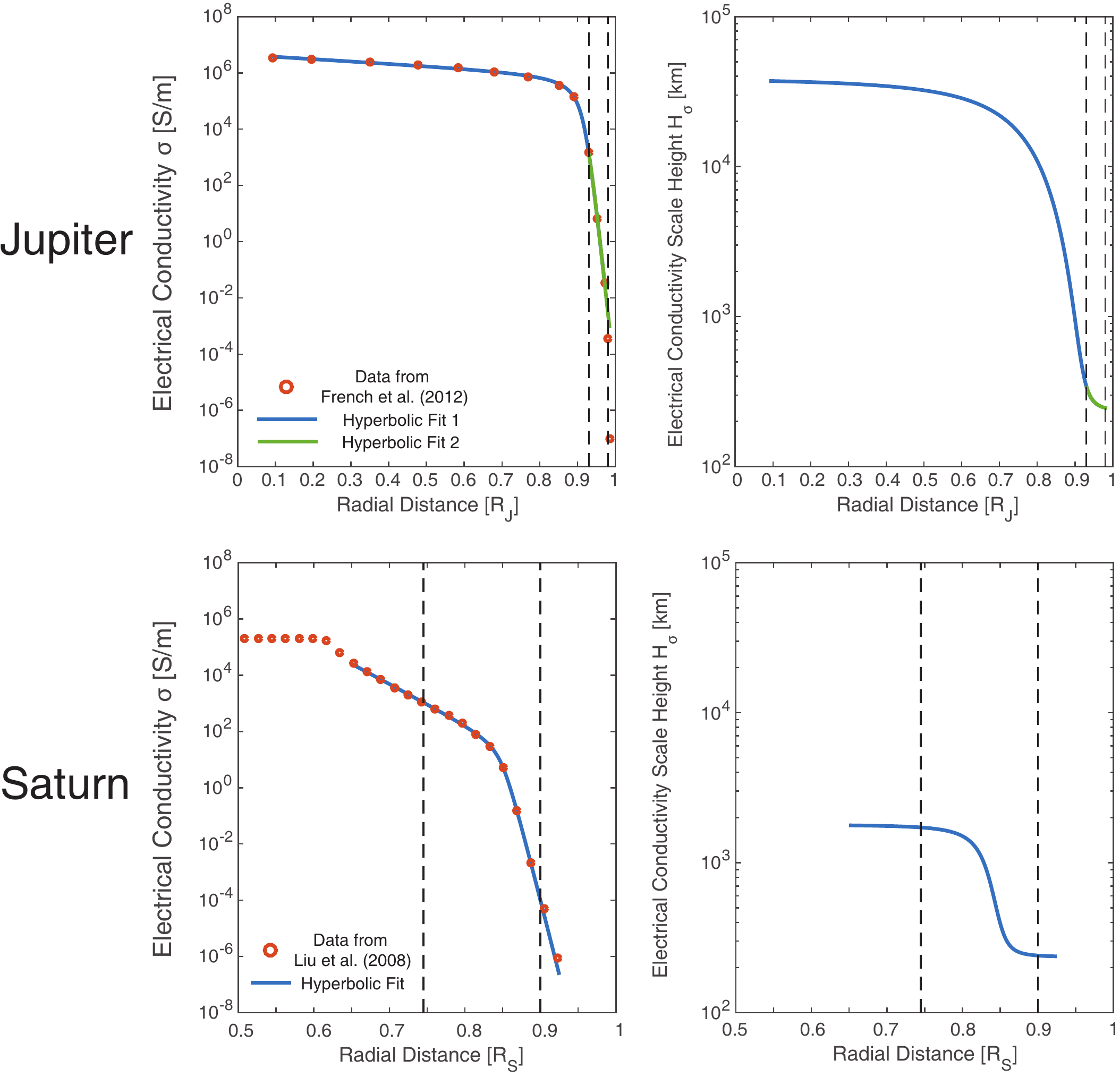}
 \caption{Electrical conductivity and the associated scale height for the interiors of Jupiter and Saturn. The data from \citet[]{French2012} and \citet[]{Liu2008} and hyperbolic fits are displayed. (The functional form and values of the coefficients of the hyperbolic fits can be found in Appendix B.) Electrical conductivity on the order of 1000 $S/m$ are reached around 0.93 $R_J$ and 0.745 $R_S$, while electrical conductivity on the order of $10^{-4}$ $S/m$ are reached around 0.98 $R_J$ and 0.90 $R_S$ }
 \label{fig:fig3}
\end{figure}

The profiles of the electrical conductivity, $\sigma$, and the associated scale-height, $H_\sigma=|\sigma/\frac{d\sigma}{dr}|$, for the interiors of Jupiter and Saturn are shown in Fig. \ref{fig:fig3}. The data from \citet[]{French2012} and \citet[]{Liu2008} and hyperbolic fits are displayed. (The functional form and values of the coefficients of the hyperbolic fits can be found in Appendix C.) Electrical conductivity on the order of 1000 $S/m$ are reached around 0.93 $R_J$ and 0.745 $R_S$ respectively. Given that the measured jovian magnetic secular variation is on the order of 1 $cm/s$ \citep[]{Yu2010, RH2016}, region with electrical conductivity much greater than 1000 $S/m$ inside Jupiter likely features zonal flows slower than 1 $cm/s$, since the magnetic Reynolds number ($Rm=UH_\sigma/\eta$) associated with 1 $cm/s$ flows there would be greater than 10. More importantly, mean-field electrodynamics becomes questionable in regions with electrical conductivity much greater than 1000 $S/m$ due to the large $Rm$ associated with the small-scale convection. Assuming a convective velocity on the order of 1 $mm/s$, and a convective length-scale on the order of $10^{-2}$ of the planetary radius \citep[]{SJ2002}, the magnetic Reynolds number associated with small-scale convection would exceed unity for regions with electrical conductivity greater than 1000 $S/m$. With the same estimate about the typical velocity and the typical length-scale of the convection, magnetic diffusivity in regions with electrical conductivity smaller than 1000 $S/m$ would dominate the (total) effective diffusivity ($\eta_E \sim \eta \gg \beta$). From these estimations, the lower boundary of the semi-conducting region of Jupiter and Saturn can be placed at 0.93 $R_J$ and 0.745 $R_S$ respectively. 


The upper boundary of the semi-conducting region can be placed at where the interaction between zonal flow and magnetic field becomes negligible. Electrical conductivity on the order of 0.01 $S/m$ is reached around 0.972 $R_J$ and 0.875 $R_S$, while electrical conductivity on the order of $10^{-4}$ $S/m$ is reached around 0.98 $R_J$ and 0.90 $R_S$. The magnetic Reynolds number associated with 100 $m/s$ flows are smaller than 0.25 at a depth with $\sigma=0.01 S/m$ and $H_\sigma=200 km$, and are smaller than $2.5\times10^{-3}$ at a depth with $\sigma=10^{-4} S/m$ and $H_\sigma=200 km$. Setting the outer boundary of the semi-conducting region at 0.98 $R_J$ and 0.90 $R_S$ would suffice for investigating the interaction between zonal flow and magnetic field under most circumstances.

We notice that the radii correspond to 0.01 $S/m$ for Jupiter and Saturn (0.972 $R_J$ and 0.875 $R_S$) turn out to be very close to the cylindrical radii at which the equatorial super-rotation maps to the deep interior along cylinders parallel to the spin-axis. The equatorial super-rotation observed at Jupiter and Saturn could extend to the deep interior with constant velocity along the spin-axis. For this particular scenario, high-degree gravity moments will be dominated by the mass redistribution associated with the equatorial super-rotation \citep[]{Liu2013, Liu2014}. 


\subsection{Free Parameters in the Calculation}

For a given zonal flow profile, a given electrical conductivity profile, and a given background magnetic field profile, the toroidal magnetic field can be uniquely determined from equation (\ref{eqn:B_Final}) without any further assumption or free-parameter. To determine the measurable poloidal magnetic perturbations, however, one needs to estimate the amplitude and profile of the dynamo $\alpha-$effect which is a big unknown. Here we make the following assumptions about the dynamo $\alpha-$effect for Jupiter and Saturn, based on our understanding of rapidly rotating convection. 1) The dynamo $\alpha-$effect is antisymmetric about the equator. 2) In each hemisphere, the statistical properties of the turbulent convection is uniform in the semi-conducting region. The dynamo $\alpha-$effect thus is inversely proportional to the magnetic diffusivity. 3) The amplitude of the dynamo $\alpha-$effect at the base of the semi-conducting region is about 10\% of the convective velocity. With the estimated convective velocity of 1 $mm/s$, the amplitude of the dynamo $\alpha-$effect is about 0.1 $mm/s$ at the base of the semi-conducting region. We adopt the following functional form for the dynamo $\alpha-$effect, 
\begin{equation}
\alpha=- \alpha_0 \frac{\eta_0}{\eta} \mathit{erf} \left( \frac{\theta-\frac{\pi}{2}}{0.005\pi}\right),
\end{equation}
where $\alpha_0$ is the amplitude of the dynamo $\alpha-$effect at the base of the semi-conducting region, $\eta_0$ is the magnetic diffusivity at the base of the semi-conducting region, $\theta$ is the co-latitude, and $erf$ is the error function. 
This functional form ensures $\alpha=-\alpha_0 \eta_0/\eta$ in the majority of the northern hemisphere, and $\alpha=\alpha_0 \eta_0/\eta$ in the majority of the southern hemisphere. It should be noted that the results to be presented do not depend on the details of the functional form. A dynamo $\alpha-$effect with a simple sine dependence on latitude and inversely proportional to magnetic diffusivity would yield very similar results. 

\subsection{An Order-of-Magnitude Analysis of the Magnetic Perturbations}

From equations (\ref{eqn:A_Final}) and (\ref{eqn:B_Final}), it is straightforward to make an order-of-magnitude analysis of the magnitude of the wind induced magnetic perturbations. It can be shown that
\begin{equation}
B_{Tor}^{Wind} \sim Rm(U_\phi) B_0,
\end{equation}
\begin{equation}
B_{Pol}^{Wind} \sim Rm(\alpha) B_{Tor},
\end{equation}
\begin{equation}
B_{Pol}^{Wind} \sim Rm(\alpha) Rm(U_\phi) B_0,
\label{eqn:BpolAmp}
\end{equation}
here $B_0$ is the amplitude of the background magnetic field, and
\begin{equation}
Rm(U_\phi)=\frac{U_\phi H_\sigma}{\eta},
\end{equation}
\begin{equation}
Rm(\alpha)=\frac{\alpha H_\sigma}{\eta},
\end{equation}
It is important to realize that the two magnetic Reynolds numbers are generally evaluated at different conductivity levels, since diffusion is efficient in the semi-conducting region. We will compare the results from numerical calculations to these scalings.

\section{Calculation for Jupiter and Saturn}

We conducted a series of calculations of zonal flow magnetic field interactions for Jupiter and Saturn. We solved the spectra representation of equations (\ref{eqn:A_Final}) and (\ref{eqn:B_Final}) using the Chebyshev collocation method (see Appendix A). The typical resolution adopted in our calculations are 480 Chebyshev grid points in the radial direction and 360 Gaussian-quadrature grid points in the latitudinal direction. Only hemispherically symmetric winds are considered for simplicity. For each planet, two surface wind profiles are built via mirroring the northern hemisphere wind to the southern hemisphere and vice versa. These surface wind profiles are first projected onto order-1 associated Legendre polynomials, $P_n^1(\cos \theta)$, then truncated at degree 100 to ensure smoothness. 

\subsection{A Single Equatorial Jet}

First, we investigate the interaction between a single equatorial zonal jet and the deep dynamo generated magnetic field. For this calculation, we are mostly interested in the case where the equatorial jet closely resembles that observed at the surface of Jupiter and Saturn. We project the observed equatorial jet along the direction of spin-axis into the deep interiors of Jupiter and Saturn, and calculated the magnetic Reynolds number associated with these flows in the equatorial plane (Fig. 4). It can be seen that the peak magnetic Reynolds number reach $\sim 0.02$ and $\sim 0.03$ for Jupiter and Saturn respectively. Relatively smooth equatorial zonal wind profiles that closely resemble those observed are adopted in the numerical calculations (Fig. 4). For these two particular calculations, we extend the outer boundary of the semi-conducting region to 0.985 $R_J$ and 0.95 $R_S$ respectively.

Fig. 5 shows some of the details for the Jupiter calculation. Panel (a) shows the zonal flows, panel (b) shows the magnetic Reynolds number associated with the zonal flows, panel (c) shows the dimensionless interaction (forcing) parameter, $s (\mathbf{B_0^*} \cdot \nabla \omega) H_\sigma^2/\eta$, here $\mathbf{B_0^*}$ is the dimensionless background magnetic field while all other quantities are dimensional, and panel (d) show the resulted toroidal magnetic field, $\Delta B_\varphi$, due to the flow shear ($\omega-$effect). The latitude radial distance projection is adopted for better visualization.

\begin{figure}[h!]
 \centering
     \includegraphics[width=0.95\textwidth]{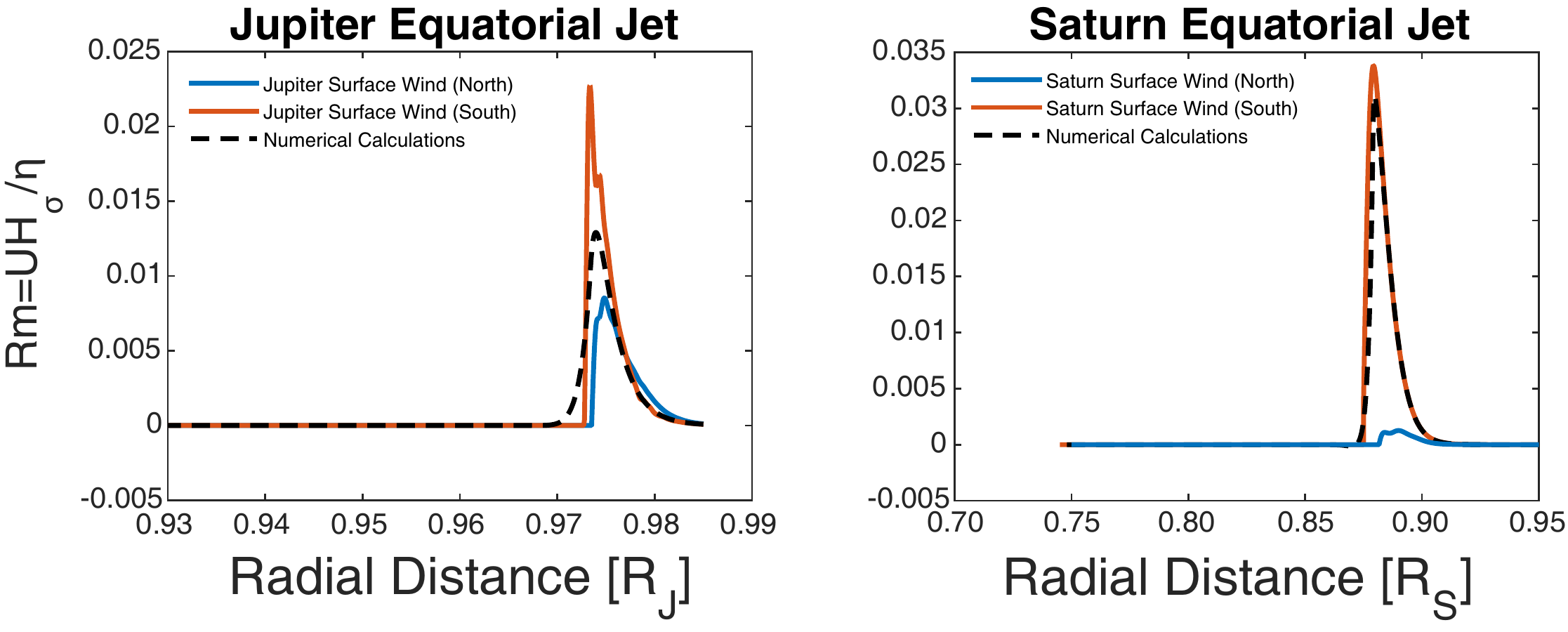}
 \caption{The magnetic Reynolds number in the equatorial plane associated with the observed surface equatorial super rotation at Jupiter and Saturn projected along the direction of the spin axis. It can be seen that the peak magnetic Reynolds number reach $\sim 0.02$ and $\sim 0.03$ for Jupiter and Saturn respectively. Relatively smooth equatorial zonal wind profiles adopted in the numerical calculations are displayed as well.}
 \label{fig:fig4}
\end{figure}

\begin{figure}[h!]
 \centering
     \includegraphics[width=0.95\textwidth]{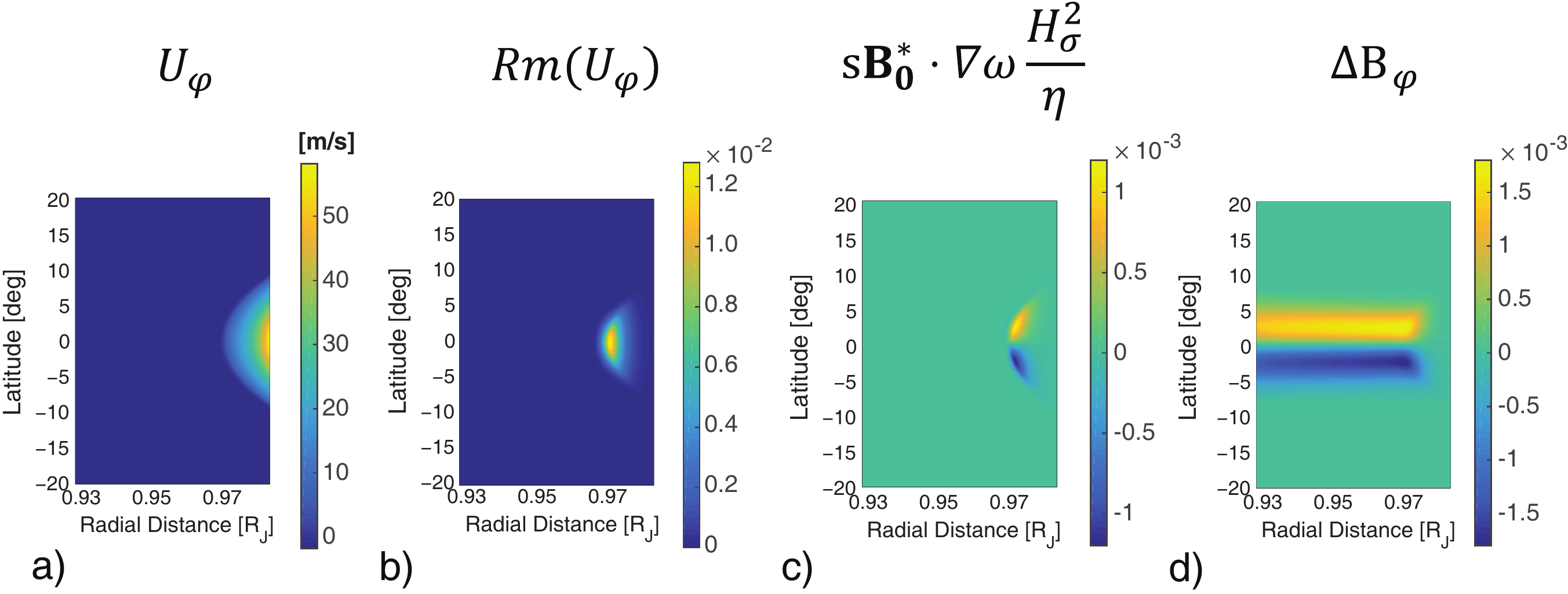}
 \caption{Physics of zonal flow magnetic field interaction for the observed equatorial jet at Jupiter. Panel (a) shows the zonal flows, panel (b) shows the magnetic Reynolds number associated with the zonal flows, panel (c) shows the dimensionless interaction (forcing) parameter, $s (\mathbf{B_0^*} \cdot \nabla \omega) H_\sigma^2/\eta$, and panel (d) shows the resulted toroidal magnetic field, $\Delta B_\varphi$, due to the shear ($\omega-$effect).}
 \label{fig:fig5}
\end{figure}

\begin{figure}[h!]
 \centering
     \includegraphics[width=0.95\textwidth]{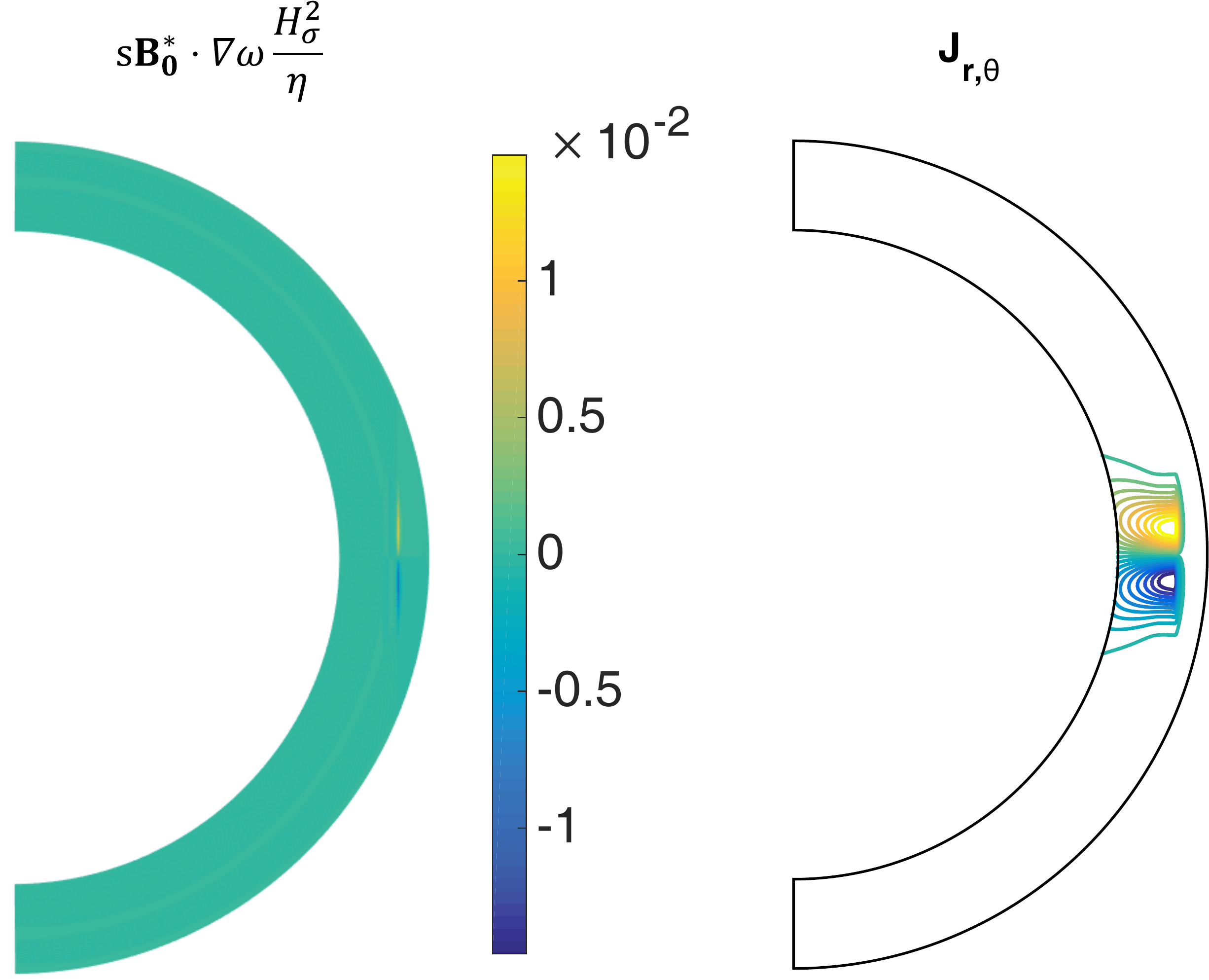}
 \caption{The dimensionless interaction (forcing) parameter, $s (\mathbf{B_0^*} \cdot \nabla \omega) H_\sigma^2/\eta$, and the stream line of the electric currents in the meridional plane in the single equatorial jet Saturn case. It can be seen that horizontal currents in the interaction region converge into radial currents and penetrate downward into regions with high electric conductivity.}
 \label{fig:Ohmic}
\end{figure}

\begin{figure}[h!]
 \centering
     \includegraphics[width=0.95\textwidth]{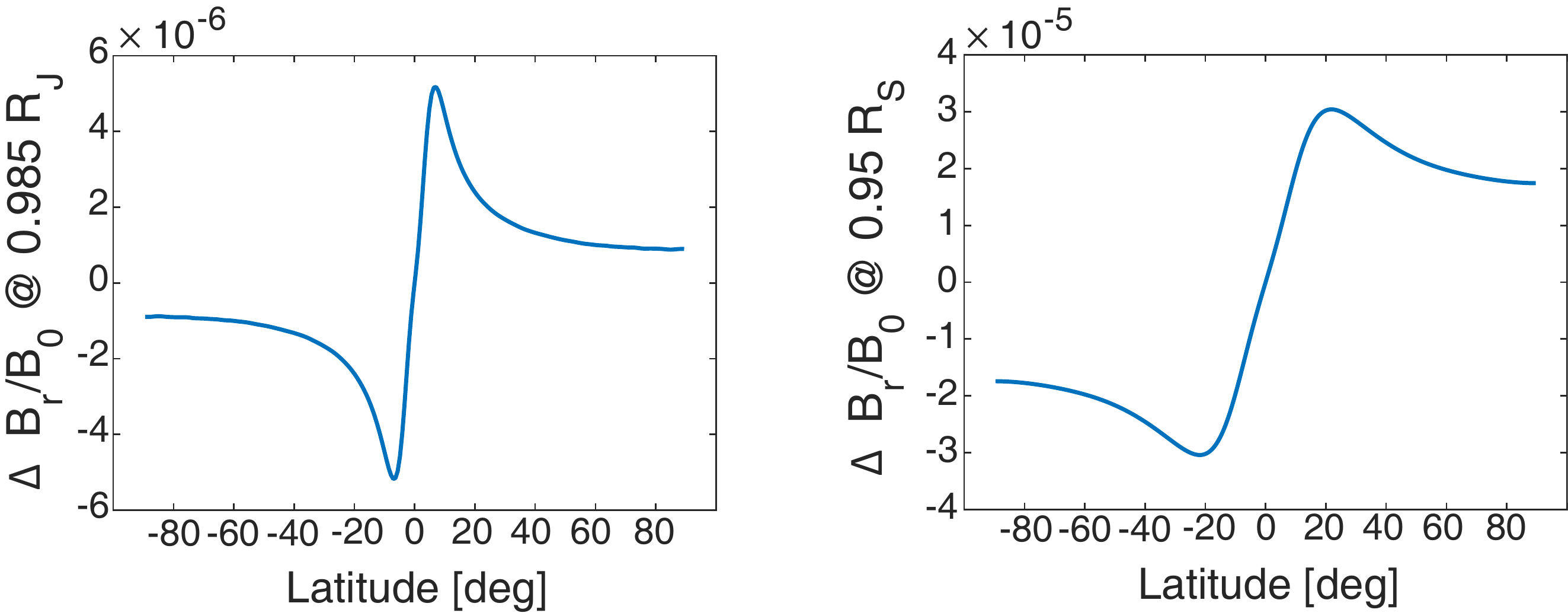}
 \caption{Profile of the wind induced $B_r$ in the single equatorial jet cases evaluated at 0.985 $R_J$ and 0.95 $R_S$ respectively.}
 \label{fig:JupInt}
\end{figure}

It can be seen that the wind induced toroidal magnetic field is on the order of $1.5 \times 10^{-3}$ of the background dipole field for Jupiter, which is about one order of magnitude lower than $Rm(U_\phi)$. This is likely due to the geometrical properties of an axial dipole magnetic field near the equator: the cylindrical radial component of a dipole magnetic field approaches zero as one approaches the equatorial plane, and yet the shear of an equatorial jet is entirely in the cylindrical radial direction. As the equatorial jet of Jupiter is very narrow, the geometrical effect is pronounced. This geometrical property is well reflected in the dimensionless interaction parameter $s (\mathbf{B_0^*} \cdot \nabla \omega) H_\sigma^2/\eta$. It can be seen in Fig. 5 that the peak toroidal field strength is very close to the peak value of the dimensionless interaction parameter.

Downward diffusion of the toroidal magnetic field is prominent in these two calculations. It is clear from Fig. 5 that although the interaction between the zonal flow and the background magnetic field peak strongly near the outer boundary, the toroidal magnetic field generated by the interaction diffuses downward. This downward diffusion can be most easily understood through visualizing the electric currents. The right panel of Fig. 6 displays the streamline of the electric current in the meridional plane for the Saturn case. It can be seen that horizontal currents in the interaction region converge into radial currents and penetrate downward into regions with high electric conductivity. Although the electrical currents penetrate into deeper regions, the Ohmic dissipation is dominated by the relatively shallow regions since $q=J^2/\sigma$. Thus for relatively constant current density as a function of radial distances, the Ohmic dissipation is dominated by regions with low electric conductivity. The integrated Ohmic dissipation are $1.5 \times 10^{-5}$ $W/m^2$ for the Jupiter case and $5 \times 10^{-6}$ $W/m^2$ for the Saturn case, which are many orders of magnitude smaller than the observed heat flux at the surface of Jupiter and Saturn.  

Fig. 7 shows the wind induced radial magnetic field at 0.985 $R_J$ and 0.95 $R_S$ in these two calculations. The induced radial magnetic field associated with the observed equatorial super rotation take a dipolar geometry and very small values: $5 \times 10^{-6}$ and $3 \times 10^{-5}$ of the background dipole field respectively. These translate into $\sim$ 2 $nT$ and $\sim$ 0.7 $nT$ for Jupiter and Saturn respectively, which are extremely unlikely to be detectable.   

Thus even if the observed surface equatorial super rotation at Jupiter and Saturn project into the interior of the planets with constant velocity along the direction of the spin-axis, only negligible modifications to the deep dynamo generated magnetic field are expected.

\subsection{Off-Equatorial Jets}

We then proceed to calculate the interaction between the off-equatorial jets and the background magnetic field. For these calculations, different zonal flow profiles in the semi-conducting region are defined by three parameters: the transition depth, $r_T$, peak amplitude at the transition depth, $U_0$, and vertical wind scale height below the transition depth, $H_\omega$. The cylindrical radial dependence of the zonal flows is simply the surface zonal wind projected along the spin-axis. Zonal flows in the semi-conducting region can be expressed as
\begin{equation}
U_\varphi(r,\theta)=f(r)U_\varphi^{Surf}(r\sin \theta),
\end{equation}
here $f(r)$ is the radial decay function which takes the following functional form
\begin{equation}
f(r) = U_{Match} \exp \left( \frac{r-r_T}{H_\omega} \right), r \le r_T
\end{equation}
\begin{equation}
f(r) = \left[ 1 + \left ( U_{Match} -1 \right ) \left( \frac{r-r_P}{r_T-r_P} \right)^{D} \right], r>r_T
\end{equation}
where
\begin{equation}
U_{Match}=\frac{U_0}{max\left[U_\varphi^{Surf}(r_T\sin \theta)\right]},
\end{equation}
\begin{equation}
D=\frac{(r_P-r_T)U_{Match}}{(1-U_{Match})H_\omega},
\end{equation}
and $r_P$ is the radius of the plant. Fig. 8 displays a few examples of the radial decay function of the zonal flows in the semi-conduction region. This radial decay function ensures the smoothness of the zonal flows and allows for larger vertical scale heights outside the transition depth. In all the off-equatorial jets calculations presented here, we fix the transition depth to 0.972 $R_J$ for the Jupiter cases, and to 0.875 $R_S$ for the Saturn cases. The choices for these particular values are guided by the observational fact that these are the depth at which the equatorial super-rotation would touch the deep interior along cylinders parallel to the spin-axis. We then surveyed $U_0$ from 10 $m/s$ to 0.1 $m/s$ and $H_\omega$ from 100 $km$ to 1000 $km$. The outer boundary of the semi-conducting region for these calculations are set at 0.98 $R_J$ and 0.90 $R_S$ respectively. We extend the outer boundary to 0.985 $R_J$ and 0.95 $R_S$ for a few test cases and observe no difference in the resulting magnetic field perturbations. 

\begin{figure}[h!]
 \centering
     \includegraphics[width=0.5\textwidth]{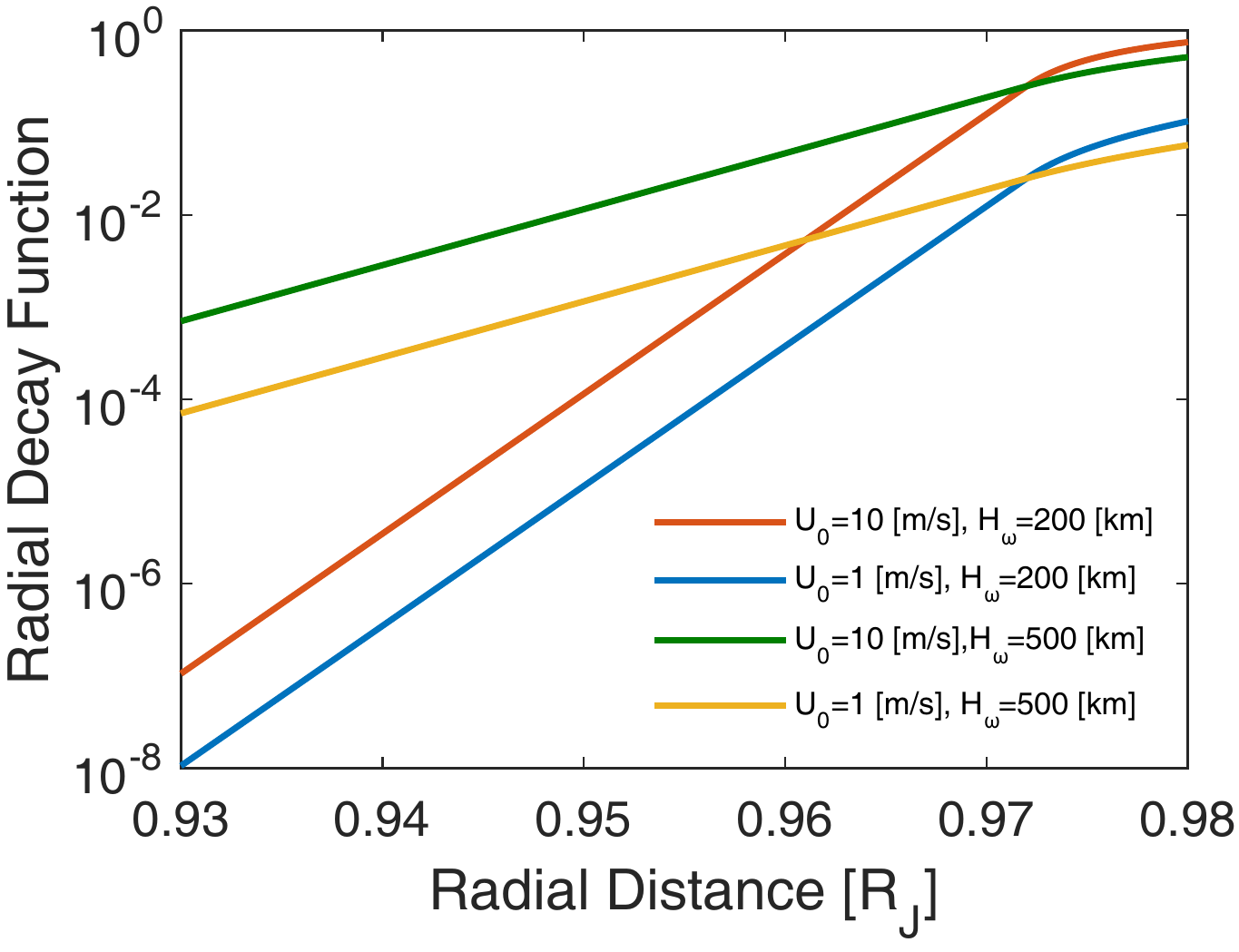}
 \caption{A few example of the radial decay function adopted for the deep zonal winds.}
 \label{fig:fig8}
\end{figure}

\begin{figure}[h!]
 \centering
     \includegraphics[width=0.95\textwidth]{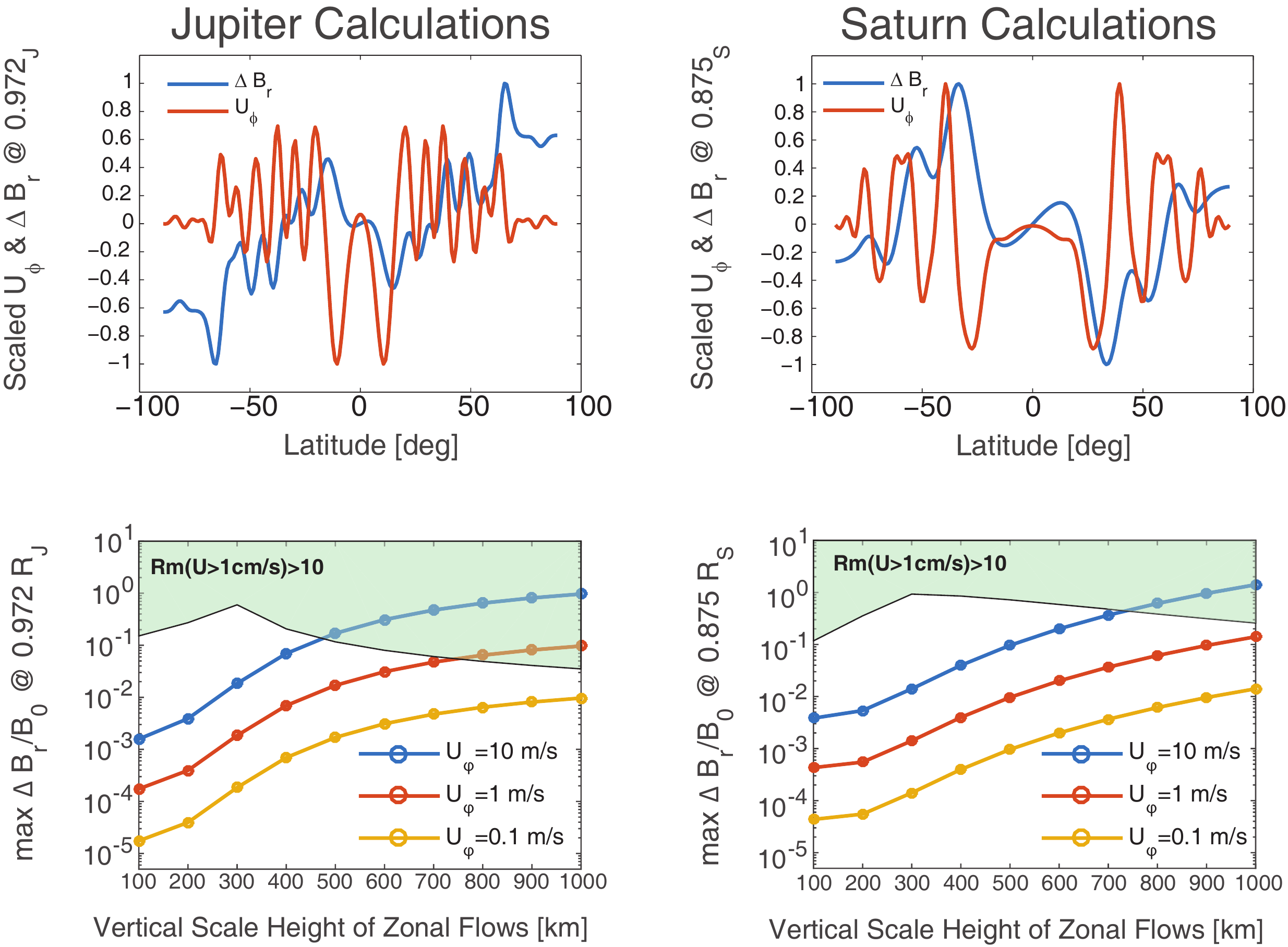}
 \caption{Profile and amplitude of wind induced magnetic field for off-equatorial jets. The upper panels show the scaled wind induced $B_r$ and the scaled zonal flow velocities at the transition depth from one calculation for Jupiter and Saturn respectively. In both cases, $U_0=1$ $m/s$ and $H_\omega=200$ $km$. It can be seen that the wind induced $B_r$ are spatially correlated with the zonal flows. The lower panels display the peak amplitude of the wind induced $B_r$ scaled to $B_0$ at the transition depth from the survey calculations. It can be seen that 1 $m/s$ wind with vertical scale height between 100 $km$ and 500 $km$ would generate poloidal magnetic perturbations on the order of 0.01\% - 1\% of the background dipole field. In the lower panels, parameter space in which magnetic Reynolds number associated with 1 $cm/s$ zonal flows exceeding 10 are shaded in light green.}
 \label{fig:fig4}
\end{figure}

The upper panels of Fig. 9 show the scaled wind induced $B_r$ and the scaled zonal flow velocities at the transition depth from one calculation for Jupiter and Saturn respectively. In both cases, $U_0=1$ $m/s$ and $H_\omega=200$ $km$. It can be seen that the magnetic perturbation generated by the zonal flow magnetic field interaction are spatially correlated with the zonal flows, even when evaluated at the transition depth. The peak amplitude of wind induced $B_r$ scaled to $B_0$ at the transition depth from the survey calculations are shown in the lower panels of Fig. 8. It can be seen that 1 $m/s$ wind with vertical scale height between 100 $km$ and 500 $km$ would generate poloidal magnetic perturbations on the order of 0.01\% - 1\% of the background dipole field when evaluated at 0.972 $R_J$ and 0.875 $R_S$. The integrated Ohmic dissipation associted with all wind profiles considered here are smaller than the observed surface heat flux of Jupiter (5 $W/m^2$) and Saturn (2 $W/m^2$). Parameter space with $Rm(U_\phi$ $>$ 1 $cm/s$) exceeding 10 are shaded in light green on the same figure.

\begin{figure}[h!]
 \centering
     \includegraphics[width=0.95\textwidth]{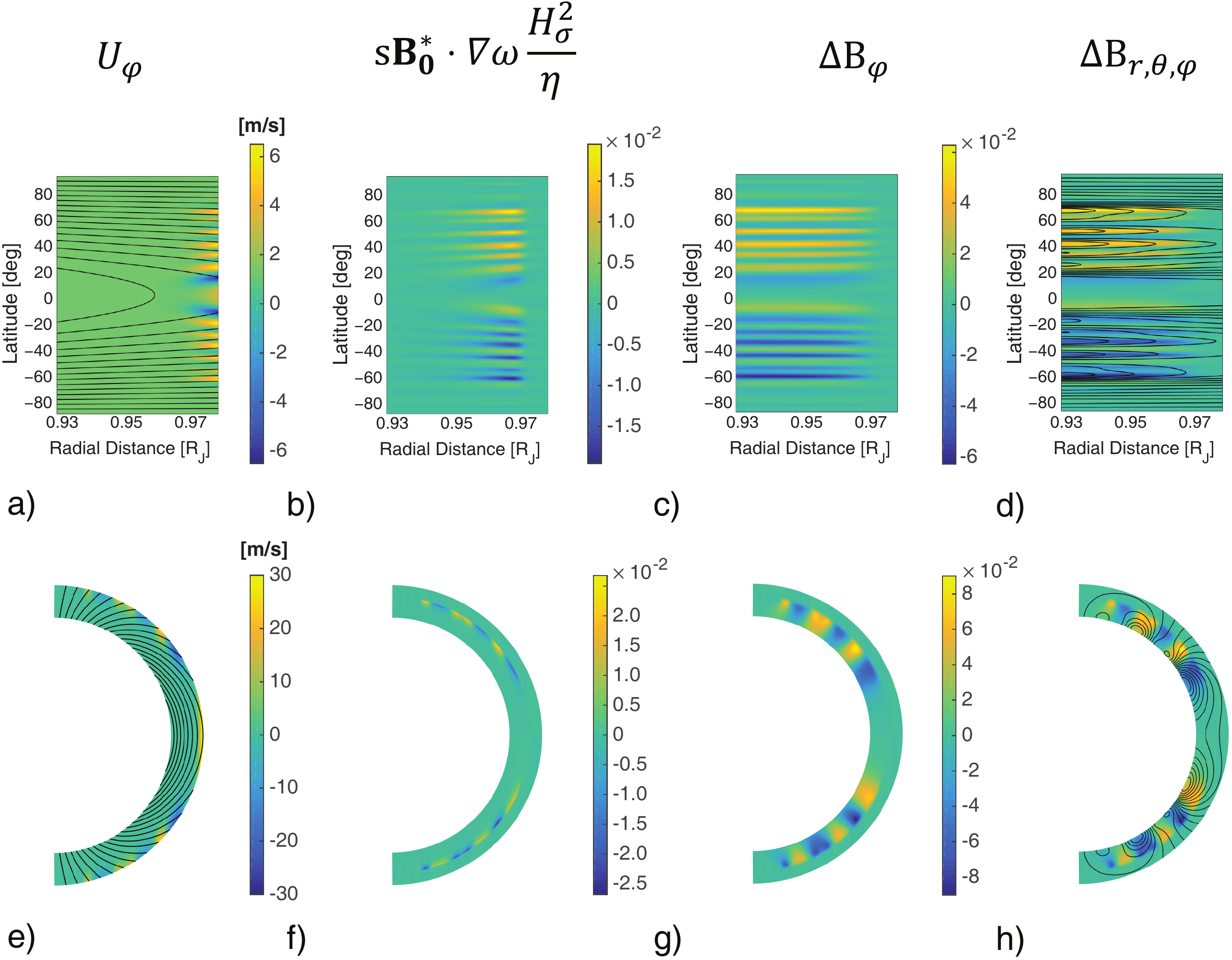}
 \caption{Physics of zonal flow magnetic field interaction for the off-equatorial jet cases. Panels (a) \& (e) show the zonal flows and the background magnetic field, panel (b) \& (f) show the dimensionless interaction (forcing) parameter, $s (\mathbf{B_0^*} \cdot \nabla \omega) H_\sigma^2/\eta$, panels (c) \& (g) show the resulted toroidal magnetic field ($\Delta B_\varphi$) due to the shear (dynamo $\omega-$effect) in colors, and panels (d) \& (h) show the resulted poloidal magnetic field ($\Delta B_{r,\theta}$) due to the dynamo $\alpha-$effect in field lines.}
 \label{fig:SatInt}
\end{figure}

Fig. 10 shows some details for the two calculations shown in the upper panels of Fig. 9. Panels (a) \& (e) show the zonal flows and the background magnetic field, panel (b) \& (f) show the dimensionless interaction (forcing) parameter, $s (\mathbf{B_0^*} \cdot \nabla \omega) H_\sigma^2/\eta$, panels (c) \& (g) show the resulted toroidal magnetic field ($\Delta B_\varphi$) due to the shear (dynamo $\omega-$effect) in colors, and panels (d) \& (h) show the resulted poloidal magnetic field ($\Delta B_{r,\theta}$) due to the dynamo $\alpha-$effect in field lines. The efficient downward diffusion of the toroidal magnetic are clearly visible in this figure. 

\begin{figure}[h!]
 \centering
     \includegraphics[width=0.95\textwidth]{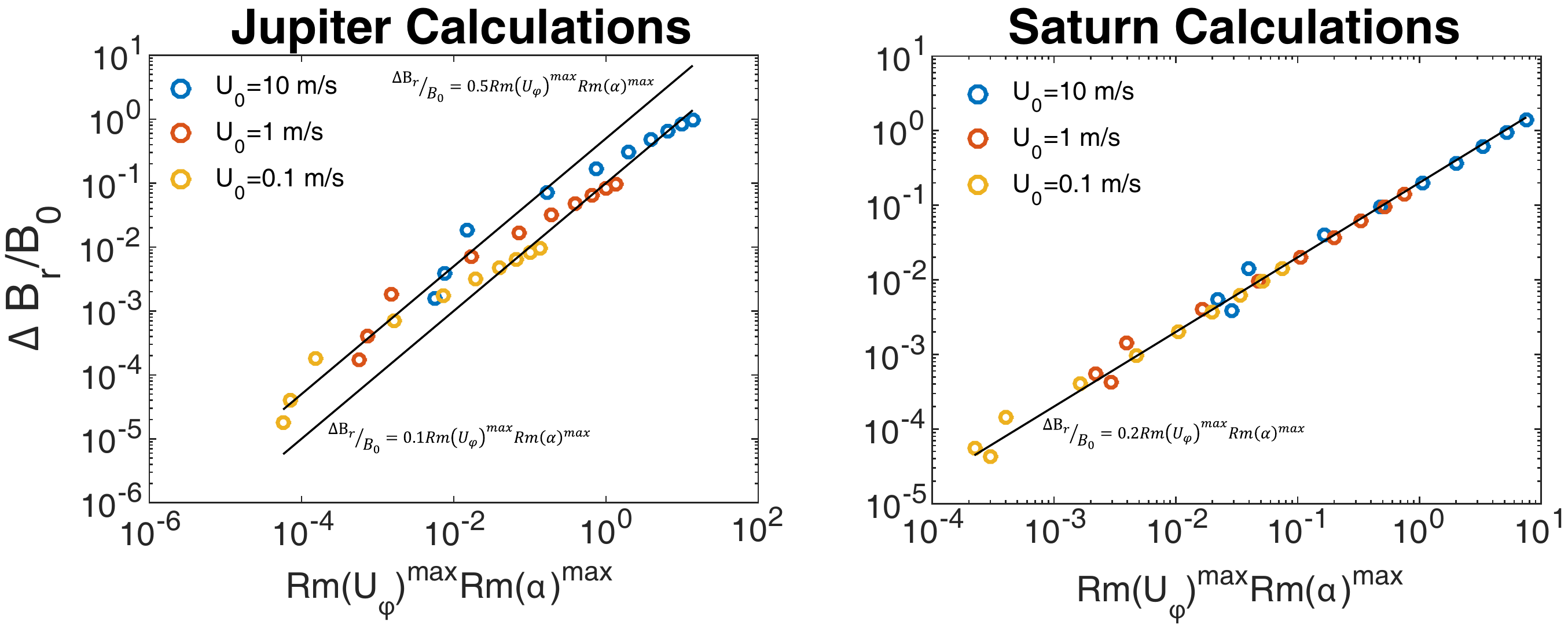}
 \caption{Amplitude of wind induced poloidal magnetic perturbations in the numerical calculations compared to the scaling relation (\ref{eqn:BpolAmp}). Almost all the Saturn calculations can be described by the scaling relation (\ref{eqn:BpolAmp}) with a numerical pre-factor of 0.2. The Jupiter calculations can be described by the scaling relation (\ref{eqn:BpolAmp}) with a numerical pre-factor between 0.1 and 0.5. The Jupiter cases with magnetic perturbations below 1\% are better described with a numerical per-factor of 0.5 while the Jupiter cases with magnetic perturbations above 1\% are better described with a numerical per-factor of 0.1.}
 \label{fig:dBrRm}
\end{figure}

Fig. 11 compares the amplitude of the wind induced poloidal magnetic field in the numerical calculations with the scaling relation (\ref{eqn:BpolAmp}). It can be seen that the scaling relation (\ref{eqn:BpolAmp}) predicts the amplitude of the poloidal perturbations reasonably well. Almost all the Saturn calculations can be described by the scaling relation (\ref{eqn:BpolAmp}) with a numerical pre-factor of 0.2. The Jupiter calculations can be described by the scaling relation (\ref{eqn:BpolAmp}) with a numerical pre-factor between 0.1 and 0.5. The Jupiter cases with magnetic perturbations below 1\% are better described with a numerical per-factor of 0.5 while the Jupiter cases with magnetic perturbations above 1\% are better described with a numerical per-factor of 0.1. Since $Rm(U_\phi)$ is only a crude proxy for the generation of toroidal magnetic field from the interaction between zonal flow and background magnetic field as discussed in section 5.1, one should not expect the scaling relation to apply with a universal pre-factor. 

These calculations indicate that if zonal flow on the order of 1 $m/s$ exist in the semi-conducting region of Jupiter and Saturn, poloidal magnetic perturbations, spatially correlated with the zonal flows, on the order of 0.01\% - 1\% of the background dipole field will be induced. These magnetic perturbations should be detectable with low altitude orbital magnetometer measurements with good latitudinal coverage, such as those to be provided by the Juno mission and the Cassini Grand Finale.

\section{Observational Detection of Wind Induced Magnetic Perturbations}

In terms of observational detection of wind induced magnetic perturbations, there are two choices to evaluate the signal: in real space and in spectral space. Given that we do not have a predictive theory for the magnetic spectra of the deep dynamo field, our analysis and conclusion concerning the detection of wind induced magnetic perturbations in spectral space should be regarded as tentative. 

Viewing the wind induced magnetic field in real space, it can be seen in the upper panels of Fig. 9 that wind induced $B_r$ have the same number of peaks as the off-equatorial zonal flows. Both the off-equatorial zonal flows and the wind induced $B_r$ in the Jupiter case have five broad peaks. And the off-equatorial zonal flows and the wind induced $B_r$ in the Saturn case both have three broad peaks. Observational detection of wind induced magnetic field in real space likely proceeds with a regional inversion of magnetic field (e.g. $B_r$ between 10 degrees latitude and 60 degrees latitude in the northern hemisphere) at 0.972 $R_J$ and 0.875 $R_S$. One would then high-pass filter the regional magnetic field and retaining the magnetic field with length-scale comparable to or shorter than the typical length-scale of the zonal winds in the latitudinal direction. This procedure would resemble the detection of regional crustal magnetic field as had been applied to Mars and Mercury \citep[e.g.][]{Johnson2015, PS2015}. The details would need to be worked out with the actual orbital trajectories of the measurements. 

\begin{figure}[h!]
 \centering
     \includegraphics[width=0.95\textwidth]{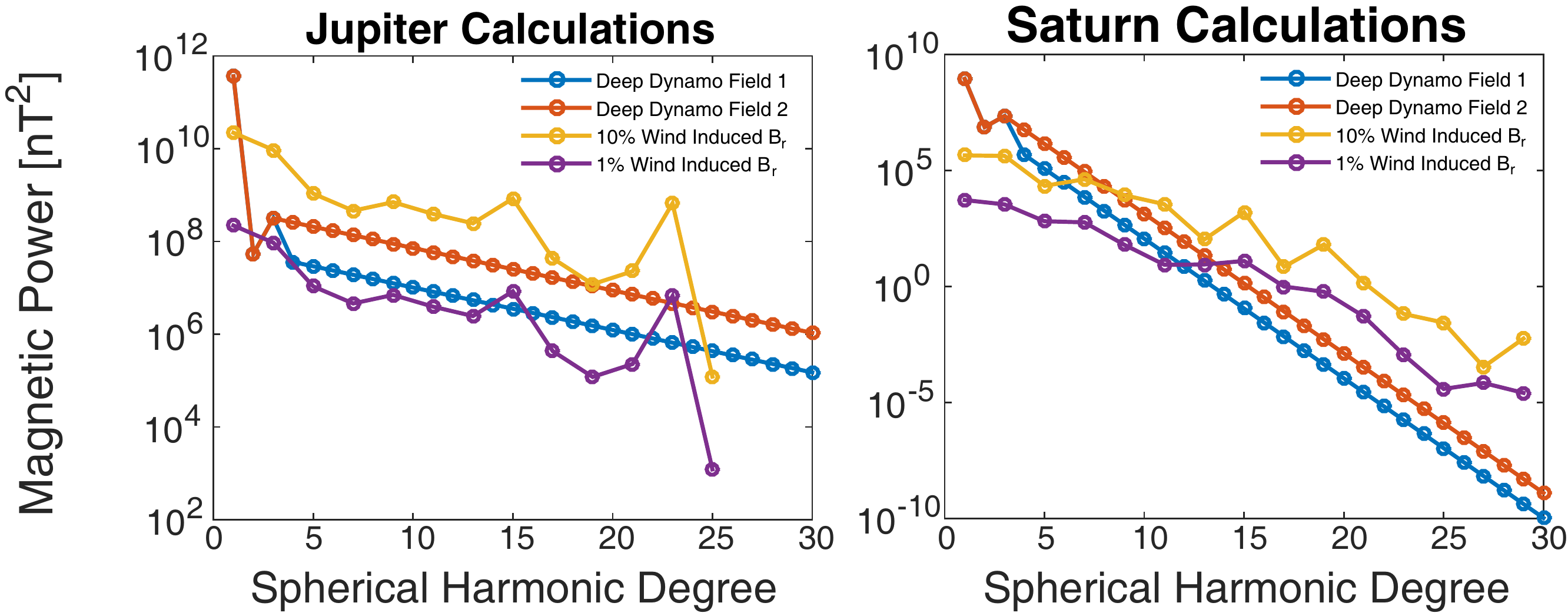}
 \caption{Magnetic power spectra of the wind induced poloidal magnetic field evaluated at the surface of Jupiter and Saturn. The power spectra of the observed low degree axial magnetic field and two empirical predictions for the dynamo generated magnetic field are displayed as well. The magnetic power from a 1 -- 10\% wind induced magnetic field will exceed that from the deep dynamo starting around degree 10 for Saturn, while magnetic power spectra of the wind induced magnetic field at Jupiter would have a similar slope as that of the deep dynamo generated magnetic field. Thus, analyzing the data in real space might be preferred for Jupiter. Given that we do not have a predictive theory for the magnetic spectra of the deep dynamo field, our analysis and conclusion concerning the detection of wind induced magnetic perturbations in spectral space should be regarded as tentative.}
 \label{fig:MagPower}
\end{figure}

One could also view the wind induced magnetic field in spectral space. Fig. 12 displays the magnetic power spectra of the wind induced magnetic field at the surface of Jupiter and Saturn. The power spectra of the observed low degree axial magnetic field and two empirical predictions for the dynamo generated magnetic field are displayed. Only axial magnetic moments are taken into account in all the spectra displayed in Fig. 12. For Jupiter, the two empirical predictions for the power spectra of deep dynamo generated magnetic field result from assuming that the axial magnetic power of high harmonic degrees equal to that of axial quadrupole at 0.90 $R_J$ and to that of axial octupole at 0.90 $R_J$. For Saturn, the two empirical predictions for the power spectra of deep dynamo generated magnetic field result from assuming that the axial magnetic power of high harmonic degrees equal to that of quadrupole at 0.5 $R_S$ and to that of octupole at 0.5 $R_S$. It can be seen from this exercise that the wind induced magnetic field may show up in the spectral space at Saturn. The magnetic power from a 1 -- 10\% wind induced magnetic field will exceed that from the deep dynamo starting around degree 10 \textbf{at} Saturn. The power spectra of the wind induced magnetic field at Jupiter would have a similar slope as that of the deep dynamo generated magnetic field. Thus, analyzing the data in real space might be preferred for Jupiter. 

\section{Summary and Conclusion}

Understanding the specific realization of hydrodynamics and magnetohydrodynamics under the conditions of giant planet interiors remain a theoretical and observational challenge. One specific aspect of the puzzle concerns the nature of the east-west dominant zonal flows observed on the surface of all four solar system giant planets. It is yet to be decided whether these flows are shallow atmospheric phenomenon or surface manifestation of deep interior dynamics. The upcoming gravity and magnetic field measurements from the Juno mission and the Cassini Grand Finale would provide observational constraints on this problem.

The physics and application of gravitational sounding of deep zonal flows inside giant planets have been extensively studied \citep[]{Hubbard1999, Kaspi2010, Liu2013, ZKS2015, WH2016, Kaspi2016, Galanti2017, Cao2017}. However, relatively small amount of efforts have been made to understand the physics and application of magnetic sounding of deep zonal flows. Here we investigate the interaction between zonal flow and magnetic field in the semi-conducting region of Jupiter and Saturn. The semi-conducting region here refers to regions with electrical conductivity between $10^{-4}$ $S/m$ and 1000 $S/m$, which resides around 0.95 $R_J$ for Jupiter and 0.80 $R_S$ for Saturn. Employing mean-field electrodynamics, we show that $\sim 1$ $m/s$ zonal flows in the semi-conducting region of Jupiter and Saturn can induce poloidal magnetic perturbations on the order of 0.01\% -- 1\% of the planetary dipole field. These poloidal magnetic perturbations would be spatially correlated with the zonal flows. Detection of such poloidal magnetic perturbations by the Juno mission and the Cassini Grand Finale would indicate that zonal flows on the order of 1 $m/s$ exist in the upper semi-conducting region of Jupiter and Saturn. Combined analysis of gravity and magnetic field would further constrain the details of deep zonal flows inside giant planets. 


\appendix

\section{Spectral Representation of the Mean Field Equations}

Given that $(\nabla^2 - \frac{1}{s^2})$ is the action in the mean-field equation under axisymmetry in the spherical coordinate (eqns 14-19), the natural functional bases to express $A$ \& $B$ are \textbf{order-1 associated legendre polynomails} $P_n^1(\cos \theta)$. Since
\begin{equation}
(\nabla^2 - \frac{1}{s^2})P_n^1(\cos \theta)=-\frac{n(n+1)}{r^2} P_n^1(\cos \theta).
\end{equation}
It should be emphasized that order-1 associated legendre polynomails are only functions of $\theta$. Their association with non-axisymmetric moments of spherical harmonics are implemented through a further multiplication with $e^{i\phi}$

Project $A$ and $B$  onto $P_n^1$
\begin{equation}
A=\sum_{n=1}^{N_{max}} a_n(r) P_n^1(\cos \theta),
\label{eqn:SpecA}
\end{equation}
\begin{equation}
B=\sum_{n=1}^{N_{max}} b_n(r) P_n^1(\cos \theta),
\end{equation}
further, project $s\mathbf{B_0}\cdot \nabla \omega$ and $\alpha B$ onto $P_n^1$
\begin{equation}
s \mathbf{B_0}\cdot \nabla \omega=\sum_{n=1}^{N_{max}} w_n(r) P_n^1(\cos \theta),
\end{equation}
\begin{equation}
\alpha B=\sum_{n=1}^{N_{max}} z_n(r) P_n^1(\cos \theta),
\end{equation}
one only needs to solve the following ordinary differential equations (ODEs) for $a_n$ \& $b_n$,
\begin{equation}
\frac{d^2 b_n}{d r^2}+(\frac{2}{r}+\frac{1}{\eta}\frac{d\eta}{dr})\frac{db_n}{dr}+[\frac{1}{r}\frac{1}{\eta}\frac{d\eta}{dr}-n(n+1)\frac{1}{r^2}]b_n+\frac{w_n}{\eta}=0,
\label{eqn:bn_Pn}
\end{equation}
\begin{equation}
\frac{d^2 a_n}{d r^2}+\frac{2}{r}\frac{da_n}{dr}-n(n+1)\frac{1}{r^2}a_n+\frac{z_n}{\eta}=0.
\label{eqn:an_Pn}
\end{equation}
The vacuum outer boundary condition is
\begin{equation}
b_n(r_o)=0,
\end{equation} 
\begin{equation}
\frac{da_n}{dr}+\frac{n+1}{r}a_n=0,
\end{equation} 
while the finite conducting steady-state inner boundary condition is
\begin{equation}
\frac{db_n}{dr}-\frac{n}{r}b_n=0,
\end{equation} 
\begin{equation}
\frac{da_n}{dr}-\frac{n}{r}a_n=0.
\end{equation} 
We further project $a_n(r)$ and $b_n(r)$ onto Chebyshev polynomials. The spectral equations (\ref{eqn:bn_Pn}) and (\ref{eqn:an_Pn}) can then be solved using Chebyshev collocation methods \citep[e.g.][]{Peyret2003, Glatzmaier2013}. The typical resolution adopted in our calculations are 480 Chebyshev grids in the radial direction and 360 Gauss-quadrature grids in the latitudinal direction. 

\section{Time Stepping the Coupled Mean Field Equation}

We have performed time stepping of the mean field dynamo equations with and without the feedback term introduced by the poloidal magnetic perturbations $s (\nabla \times A \hat{\mathbf{e_\phi}})\cdot \nabla \omega$. Neglecting meridional circulation, the fully coupled mean-field equation with the feedback term read
\begin{equation}
\frac{\partial A}{\partial t} = \alpha B + \eta_E (\nabla^2 - \frac{1}{s^2})A,
\label{eqn:A_F2}
\end{equation}
\begin{equation}
\frac{\partial B}{\partial t} = s\mathbf{B_0}\cdot \nabla \omega + s (\nabla \times A \hat{\mathbf{e_\phi}})\cdot \nabla \omega + \eta_E (\nabla^2 - \frac{1}{s^2})B + \frac{1}{r}\frac{d\eta_E}{dr}\frac{\partial (rB)}{\partial r}.
\label{eqn:B_F2}
\end{equation}

Crank-Nicolson time integration scheme is adopted for the linear terms, while second-order Adams-Bashforth integration scheme is adopted for the non-linear terms \citep[e.g.][]{Glatzmaier2013}. Fig. B.13 compares time stepping of the fully coupled equation (\ref{eqn:A_F2} - \ref{eqn:B_F2}) to the steady-state solution of the partially de-coupled equation  (\ref{eqn:A_Final} - \ref{eqn:B_Final}). It can be seen that the wind-induced poloidal magnetic perturbations calculated from these two approaches agree within 5\% when the wind-induced perturbation is smaller than ~20\% of the background field. 

\begin{figure}[h!]
 \centering
     \includegraphics[width=0.95\textwidth]{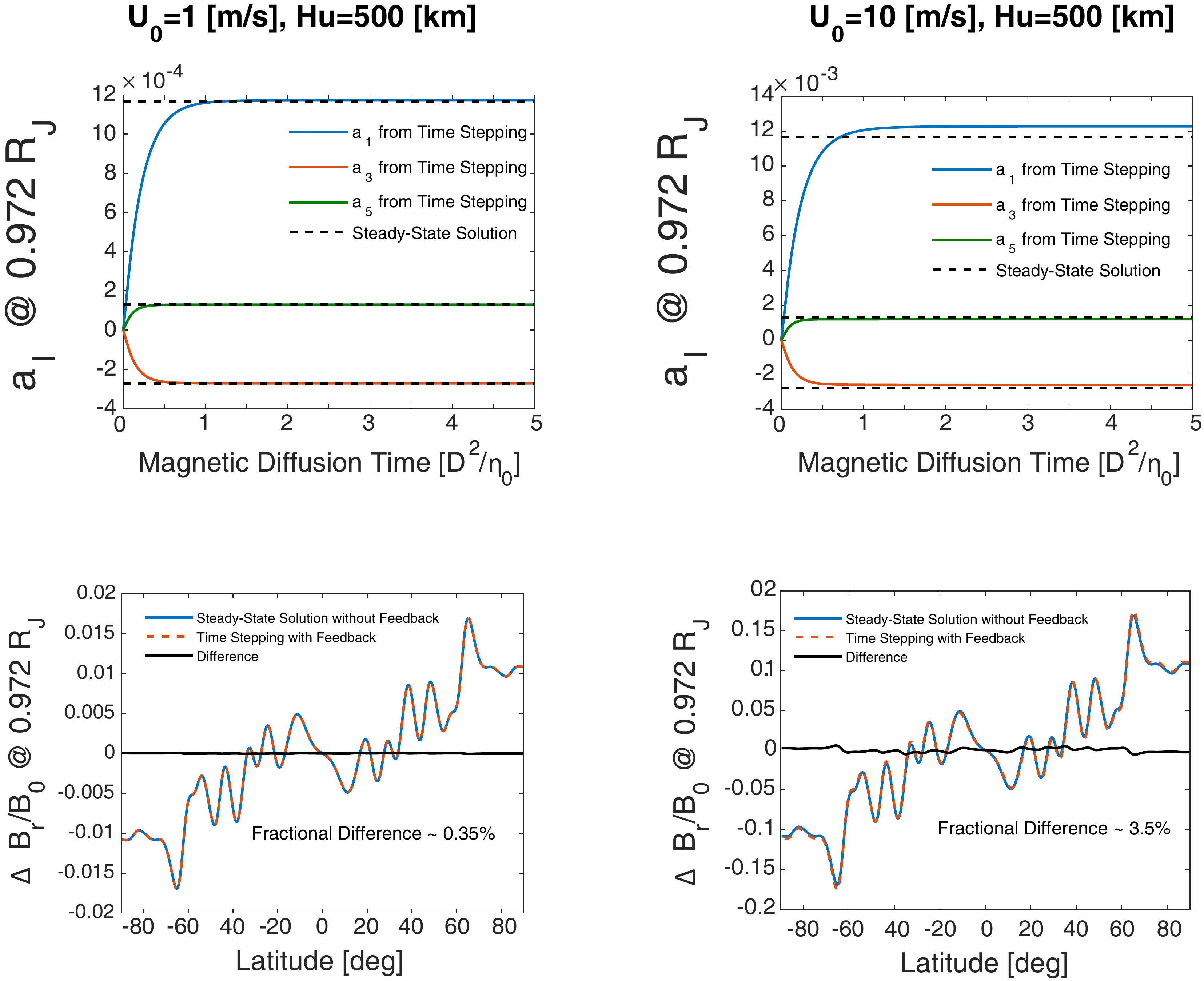}
 \caption{Comparison of wind induced poloidal magnetic perturbations from time stepping of the coupled mean-field equation (\ref{eqn:A_F2} - \ref{eqn:B_F2})  and steady-state solution of the partially decoupled mean-field equation (\ref{eqn:A_Final} - \ref{eqn:B_Final}). Upper panels show the time evolution of the solutions in Legendre space (e.g. equation \ref{eqn:SpecA}), while the lower panels show the solutions in real space. It can be seen that results from time stepping of the coupled mean-field equation agree within 5\% of the steady-state solution of the partially decoupled mean-field equation, when the wind-induced perturbation is smaller than ~20\% of the background field}
 \label{fig:TimeStepping}
\end{figure}

\section{Hyperbolic Fit to the Electrical Conductivity Profile of Jupiter and Saturn}

Given the super-exponential decay of electrical conductivity in the outer part of Jupiter and Saturn, we adopted a hyperbolic fit to ensure convergence in our numerical calculations following \citet[]{Jones2014}. The functional form for the magnetic diffusivity is
\begin{equation}
\eta=\exp\left(u+\sqrt{u^2-v}\right),
\end{equation}
\begin{equation}
u=\frac{1}{2}\left[ (g_1+g_3)\frac{r}{r_P}-g_2-g_4\right],
\end{equation}
\begin{equation}
v=\left(g_1\frac{r}{r_P}-g_2\right)\left(g_3\frac{r}{r_P}-g_4\right)-g_5,
\end{equation}
and
\begin{equation}
\left(log \eta -g_1\frac{r}{r_P}+g_2 \right) \left( log \eta -g_3\frac{r}{r_P}+g_4 \right)=g_5.
\end{equation}
The numerical values for the fits to Jupiter and Saturn are given in the following table. The hyperbolic fit to Saturn is only valid for regions between 0.65 $R_S$ and 0.90 $R_S$. The fit, \textit{Jupiter 1}, is valid for regions below 0.972 $R_J$, while the fit, \textit{Jupiter 2}, is valid for regions between 0.93 $R_J$ and 0.98 $R_J$.

\begin{table}[!h] \footnotesize
\center
\caption{Values of the Coefficients of the Hyperbolic Fit}
\begin{tabular}{ r | c | c | c | c | c }
  \hline
  \hline
     & $g_1$ & $g_2$ & $g_3$ & $g_4$ & $g_5$ \\
  \hline			
  \textit{Jupiter 1} & 299.0800 & 274.9000 & 1.7781 & 1.8010 & 20.2800 \\
  \textit{Jupiter 2} & 645.8075 & 611.4202 & 246.0274 & 222.5444 & 0.2192 \\
  \textit{Saturn} & 274.9279 & 200.6415 & 32.4004 & 17.6856 & 3.5435 \\ 
  \hline  
  \end{tabular}
\end{table}



\section{Meridional Circulation and its Transportation of Magnetic Fields}

Meridional circulation associated with deep zonal flows of giant planets have been extensively discussed in \citet[]{SchneiderL2009} and \citet[]{LiuS2010}. The general principle is the following: any stress associated with differential rotation in the system (e.g. Reynolds stress, Lorentz force, viscous stress) would drive meridional circulation. Here we derive the magnetic Reynolds number associate with the meridional circulation driven by the Lorentz force. This quantify the efficiency of meridional circulation in transporting magnetic fields. 

In the semi-conducting region, the meridional circulation driven by the Lorentz force would satisfying the following condition
\begin{equation}
2\rho \mathbf{\Omega} \times \mathbf{u_{MC}} = \frac{ \left( \nabla \times \mathbf{B} \right) \times \mathbf{B}}{\mu_0}.
\end{equation} 
The order-of-magnitude estimation of the meridional circulation thus is
\begin{equation}
u_{MC}=\frac{1}{2\rho\Omega}\frac{B^2}{\mu_0 l_J} Rm_{U_\varphi},
\label{eqn:uMC}
\end{equation} 
where $Rm_{U_\varphi}=U_\varphi H_\sigma/\eta$. The magnetic Reynolds number associated with the meridional circulation is
\begin{equation}
Rm_{MC}=\frac{u_{MC} H_\sigma}{\eta}.
\label{eqn:RmMC}
\end{equation} 
Substitute (\ref{eqn:uMC}) into (\ref{eqn:RmMC}), we get
\begin{equation}
Rm_{MC}=\frac{\sigma B^2}{2\rho \Omega}\frac{H_\sigma}{l_J}Rm_{U_\varphi},
\label{eqn:RmMCe}
\end{equation} 
in which $\sigma B^2/2\rho \Omega$ is simply the Elsasser number. 
With any reasonable assumption about $Rm_{U_\varphi}$, the magnetic Reynolds number associated with meridional circulation in the semi-conduction region of Jupiter and Saturn would be much smaller than unity. (Remember that Ohmic dissipation constraint excludes the possibility that $Rm_{U_\varphi}$ can reach 10 in the semi-conducting region of Jupiter and Saturn.) Moreover, this is a local effect. The advection of the poloidal field lines by the meridional circulation acts locally at where there are zonal wind shear.  For example, assuming $l_J=H_\sigma$, $Rm_{U_\varphi} \sim 1$ at 0.972 $R_J$ and 0.875 $R_S$ respectively, $Rm_{MC}$ would be on the order of $10^{-8}$ and $10^{-10}$. Thus the meridional circulation driven by the Lorentz force in the semi-conducting region of giant planet would be rather inefficient at transporting magnetic fields.

\bibliographystyle{elsarticle-harv}





\end{document}